\documentclass[12pt]{article}
\pdfoutput=1
\usepackage{subfigure}
\usepackage{amssymb,amsmath}
\usepackage{graphicx}
\usepackage{color}
\usepackage[colorlinks=true
,urlcolor=blue
,citecolor=blue
,linkcolor=blue
,pagecolor=blue
,linktocpage=true
,pdfproducer=medialab
]{hyperref}
\usepackage[a4paper]{geometry}

\makeatletter \renewcommand{\@dotsep}{10000} \makeatother

\newcommand{\beq}{\begin{equation}}
\newcommand{\eeq}{\end{equation}}
\newcommand{\bea}{\begin{eqnarray}}
\newcommand{\eea}{\end{eqnarray}}

\begin{document}

\begin{flushright}
OSU-HEP-14-08
\end{flushright}

\vspace*{0.2in}

\begin{center}

 {\Large\bf    Muon \boldmath{$g-2$},  125 GeV Higgs and  Neutralino Dark \\
 \vspace{0.2cm}
   Matter in sMSSM
  } \vspace{1cm}

{\large   K.S. Babu$^\dagger$\footnote{E-mail:kaladi.babu@okstate.edu},  Ilia Gogoladze$^\ast$\footnote{E-mail: ilia@bartol.udel.edu;\\
\hspace*{0.5cm} On  leave of absence from Andronikashvili Institute
of Physics,  Tbilisi, Georgia.},  Qaisar Shafi$^\ast$\footnote{ E-mail:
shafi@bartol.udel.edu} and Cem Salih$\ddot{\rm U}$n$^{\ast,\diamond}\hspace{0.05cm}$\footnote{
E-mail: cemsalihun@uludag.edu.tr}} \vspace{.5cm}

{\baselineskip 20pt \it
$^\dagger$Department of Physics, Oklahoma State University, Stillwater, OK
74078, USA \\
$^\ast$Bartol Research Institute, Department of Physics and Astronomy, \\
University of Delaware, Newark, DE 19716, USA \\
$^\diamond$ Department of Physics, Uluda\~{g} University, TR16059 Bursa, Turkey}


\vspace{1.0cm}
 {\bf Abstract}
\end{center}

We discuss the sparticle (and Higgs) spectrum in a class of flavor symmetry--based minimal supersymmetric standard models, referred to here as sMSSM.
In this framework the SUSY breaking Lagrangian takes the most general form consistent with a grand unified symmetry such as $SO(10)$ and a non-Abelian flavor symmetry acting on the three families with either a {\bf 2+1} or a {\bf 3} family assignment.  Models based on gauged $SU(2)$ and $SO(3)$ flavor symmetry, as well as non-Abelian discrete symmetries such as $S_3$ and $A_4$, have been suggested which fall into this category.  These models describe supersymmetry breaking in terms of seven phenomenological parameters.  The soft supersymmetry breaking masses at $M_{\rm GUT}$ of all sfermions of the first two families are equal in sMSSM, which differ in general from the corresponding third family mass. In such a framework we show that the muon $g-2$ anomaly, the observed Higgs boson mass of $\sim 125$ GeV, and the observed relic neutralino dark matter abundance can be simultaneously accommodated. The resolution of the muon $g-2$ anomaly in particular yields the result that the first two generation squark masses, as well the gluino mass, should be $\lesssim 2$ TeV, which will be tested at LHC14.

\newpage

\renewcommand{\thefootnote}{\arabic{footnote}}
\setcounter{footnote}{0}



\section{\label{ch:introduction}Introduction}

The ATLAS and CMS experiments at the Large Hadron Collider (LHC)  have independently reported the discovery \cite{:2012gk, :2012gu} of a  Standard Model (SM)--like Higgs boson resonance of mass $m_h \simeq 125-126$ GeV
 using the combined 7~TeV and 8~TeV data. This discovery is compatible with low scale supersymmetry,  since   the Minimal
Supersymmetric Standard Model (MSSM)  predicts an upper bound of $m_h \lesssim 135$ GeV for the lightest  CP-even Higgs boson,
if the superparticle masses are assumed to not exceed several TeV \cite{Carena:2002es}. On the other hand,  no signals have shown up for supersymmetric particles at the LHC first run, and the current lower bounds on the colored sparticle masses
\begin{equation}
  m_{\tilde{g}} \gtrsim  1.4~{\rm TeV}~ ({\rm for}~ m_{\tilde{g}}\sim m_{\tilde{q}})~~~ {\rm and}~~~
m_{\tilde{g}}\gtrsim 0.9~{\rm TeV}~ ({\rm for}~ m_{\tilde{g}}\ll
m_{\tilde{q}}) ~\cite{Aad:2012fqa,Chatrchyan:2012jx}
\label{bound1}
\end{equation}
 have created some skepticism about naturalness arguments for the Higgs mass based on low scale supersymmetry.

Although the sparticle mass bounds in Eq. (\ref{bound1}) are mostly derived for $R$-parity conserving constrained MSSM (cMSSM), they are applicable to a wider  class of low scale supersymmetric models.   There exist regions  in the MSSM parameter space where the bounds in Eq. (\ref{bound1}) can be relaxed by introducing $R$-parity violating couplings that break baryon number \cite{Bhattacherjee:2013gr}, but if the mass of the top quark superpartner, the stop, is below a TeV, the Higgs mass would be unacceptably small.  Furthermore, neutralino dark matter
will be lost in this case, owing to the violation of $R$-parity.  Low scale supersymmetry can indeed accommodate a Higgs boson with mass $m_h \simeq 125 \rm \ GeV$ in the MSSM  while preserving neutralino dark matter, but it requires either a large, ${\cal O} (\mathrm{few}-10)$ TeV,  stop  mass, or a relatively large soft supersymmetry  breaking (SSB) trilinear $A_t$-term, along with a stop  mass of around a TeV  \cite{Djouadi:2005gj}.

One of the most popular assumptions in low scale supersymmetric theories  is that of universal soft supersymmetry breaking  mass terms for the three   generations of sfermions. This assumption is mainly motivated by the constraints obtained from flavor-changing neutral currents (FCNC) processes \cite{Martin:1997ns}, with inspiration from minimal supergravity Lagrangian \cite{msugra}.  A practical outcome
of three family universality of soft masses is that it would lead to heavy sleptons in the spectrum, since the stop should be heavy to fit the Higgs boson mass.  Note, however, that the universality assumption does not follow
from any symmetry principle, and as we elaborate here, may be relaxed in a controlled fashion based on underlying symmetries.  Such a framework
is referred to here as sMSSM, for flavor symmetry-based minimal supersymmetric standard model.

 The Standard Model prediction for the anomalous magnetic moment of the muon, $a_{\mu}=(g-2)_{\mu}/2$ (muon $g-2$) \cite{Davier:2010nc}, has a discrepancy with the experimental results \cite{Bennett:2006fi}:
\begin{eqnarray}
\Delta a_{\mu}\equiv a_{\mu}({\rm exp})-a_{\mu}({\rm SM})= (28.6 \pm 8.0) \times 10^{-10}~.
\end{eqnarray}
If supersymmetry is to offer a  solution to this discrepancy, the   smuon and gaugino (bino or wino) SSB masses  should  be ${\cal O}(100)$ GeV. Thus, it is hard to simultaneously  explain the  observed Higgs boson mass and resolve the muon $g-2$ anomaly if  universality of all sfermion soft masses is imposed at the grand unified theory (GUT) scale, as in cMSSM.

Recently there have been several attempts to reconcile this (presumed) tension between muon $g-2$ and Higgs boson mass within the  MSSM framework by assuming {non-universal SSB mass terms for the gauginos \cite{Akula:2013ioa} or the sfermions \cite{Ibe:2013oha} at the GUT scale}. A simultaneous explanation of $m_h$ and muon $g-2$ is possible \cite{Ajaib:2014ana} even with  $t-b-\tau$ Yukawa coupling unification condition  \cite{yukawaUn}. It has been known for some time \cite{Baer:2004xx} that constraints from FCNC processes are very mild and easily satisfied   for the case {in which}  the third generation sfermion masses  are split from those of the first two generations. However, when the muon $g-2$ anomaly and the Higgs boson mass are simultaneously explained with non-universal gaugino and/or sfermion masses, the correct relic abundance of neutralino dark matter is typically not obtained \cite{Ibe:2013oha}. Consistency with the observed dark matter abundance would further constrain the SUSY parameter space.

In this paper we develop further the framework of flavor symmetry--based minimal supersymmetric standard model, sMSSM, suggested recently \cite{Babu:2014sga}.
It will be shown that in this framework, which consists of seven phenomenological parameters that describe supersymmetry breaking,
it is possible to explain the muon $g-2$ anomaly and the Higgs boson mass simultaneously, along with the observed dark matter abundance.
While the parameter set of sMSSM (seven) is larger
than that of cMSSM (four), it is still rather restrictive.  In comparison, the phenomenological MSSM (pMSSM) \cite{pMSSM}
describes SUSY breaking in terms of nineteen parameters.  In the sMSSM framework SUSY breaking is dictated by symmetry considerations alone.
It is realized by combining a grand unified symmetry such as $SO(10)$ with a flavor symmetry acting
on the three families which could be a gauge symmetry based on $SU(2)$ or $SO(3)$ or a discrete non-Abelian symmetry such as $S_3$ or $A_4$.  These models admit either a {\bf 2+1} or a {\bf 3} family assignment.   Both assignments would lead to the same low energy phenomenology, since a large top quark mass effectively breaks the {\bf 3} assignment down to a {\bf 2+1} assignment.  The soft SUSY breaking Lagrangian is the most general one consistent with these symmetries. FCNC processes mediated by SUSY particles are adequately suppressed by the flavor symmetry, while the grand unified symmetry further reduces the parameter set. As a consequence of these symmetries, the soft masses of the first two families are equal, which differs from that of the third family. This additional freedom helps explain the muon $g-2$, $m_h$ and dark matter abundance simultaneously. The framework is still rather restrictive, leading to the result that the sfermions of the first two families, as well as the gluino, should have masses below about 2 TeV, which will be tested in the near future at the LHC14.

The outline for the rest of the paper is as follows.
In Section \ref{sMSSM} we summarize the salient features of flavor symmetry--based MSSM.
In Section \ref{g-2} we briefly describe the dominant contributions to the muon anomalous magnetic moment {arising} from  low scale supersymmetry.
In Section \ref{constraintsSection} we summarize the scanning procedure and the
experimental constraints applied in our analysis.  Here we also  present the parameter space
that we scan over. Our results are presented in Section \ref{spectroscopySection}.  Section \ref{conclusions} has our conclusions.

\section{Flavor symmetry--based minimal supersymmetric standard model: sMSSM}

\label{sMSSM}

In this Section we provide a brief description of the sMSSM setup and its motivations \cite{Babu:2014sga}.
We also describe at the end of this section a complete model based on $SU(2)$ flavor symmetry that leads to sMSSM phenomenology at energies
below the GUT scale. We refer the reader to Ref. \cite{Babu:2014sga} for a more detailed discussion including additional models that generate the sMSSM spectrum.

In supergravity models, it is generally
assumed that supersymmetry breaks dynamically in a hidden sector, which is communicated to the visible sector via gravity.  With no further restrictions imposed, this setup would lead to over a hundred parameters in the soft SUSY breaking Lagrangian of MSSM, assuming that $R$--parity remains unbroken.  These parameters are {phenomenologically restricted by stringent constraints} from flavor changing neutral currents that the SUSY particles mediate.  To satisfy such constraints, it is often assumed that the sfermions of all three generations have a universal mass at the GUT scale.  In the constrained MSSM, for example, SUSY breaking is described by a set of four parameters, traditionally chosen to be $\{m_{0},\, M_{1/2},\, A_0,\,\tan\beta\}$, along with a discrete parameter ${\rm sgn}(\mu)$.  Such a choice, however, is not dictated by any symmetry argument, and modifications of this scheme have been widely discussed.  An example is the phenomenological MSSM (pMSSM), which describes the soft SUSY breaking
Lagrangian in terms of nineteen parameters, chosen such that SUSY mediated flavor violation is sufficiently suppressed.  As in the case of
cMSSM this setup is also not dictated by an underlying symmetry. The flavor symmetry--based minimal supersymmetric standard model  (sMSSM)
suggested in Ref. \cite{Babu:2014sga} is a framework for controlled SUSY breaking based on symmetry principles. As the framework is based on gauge symmetries, the Lagrangian is guaranteed to be protected against quantum gravitational corrections.

In the sMSSM framework the soft SUSY breaking Lagrangian is the most general one consistent with two symmetries.  First, it is compatible with a grand unified symmetry such as $SO(10)$.  Second, a non-Abelian flavor symmetry of gauge origin acts on the three families with either a {\bf 2+1} or a {\bf 3} family assignment.  This symmetry suppresses SUSY mediated flavor violation.  The grand unified symmetry, which is well motivated, and
supported by the merging of the three gauge couplings at a GUT scale of $\approx 2 \times 10^{16}$ GeV within MSSM, reduces the soft SUSY
breaking parameters considerably.  For example, gaugino mass unification is implied by GUT, which reduces the gaugino soft parameters of
MSSM from three down to one.  Similarly, all members of a family would have a common soft mass, as they are unified into a {\bf 16}-plet of
$SO(10)$.   Combined with the non-Abelian flavor symmetry, the 15 soft squared mass parameters of the 15 chiral fermions of the MSSM are reduced to just two in sMSSM. The SUSY phenomenology of sMSSM is described by seven parameters, chosen to be
\begin{equation}
\{m_{{1,2}},\, m_{{3}},\, M_{1/2},\, A_0,\,\tan\beta,\, \mu,\, m_{A}\}~.
\label{para}
\end{equation}
Here $m_{1,2}$ is the common mass of the first two family sfermions, while $m_{3}$ is the soft mass of the third family sfermions.
$M_{1/2}$ is the unified gaugino mass and $m_A$ is the mass of the pseudoscalar Higgs boson.
We shall now describe how the symmetries
of sMSSM lead to the parameter set of Eq. (\ref{para}).

A non-Abelian flavor symmetry, denoted as $H$, acts on the three families in sMSSM.  Ideally any symmetry should be a gauge symmetry,
which suggests $SU(2)$, $SO(3)$ and $SU(3)$ as possible candidates for $H$ as these groups contain {\bf 2} and {\bf 3}--dimensional
irreducible representations.  Among these, $SU(2)$ and $SO(3)$ can yield simple and realistic models of SUSY breaking and simultaneously
generate realistic fermion masses \cite{Babu:2014sga}, while this is not easily achieved in the case with $SU(3)$. Note that the representations of
$SU(2)$ and $SO(3)$ are (pseudo)real, and gauge theories based on these groups are automatically free of triangle anomalies, which
is not the case with $SU(3)$.  Gauging a flavor symmetry, however, is generally problematic in SUSY models, as it induces new and potentially dangerous flavor violation via the $D$-terms \cite{murayama}.  In Ref. \cite{Babu:2014sga} an interchange symmetry was suggested acting on a pair of doublets that break $SU(2)$ which would set the $D$-terms to zero.  Similarly, a simple solution for the $D$-term problem was found in the case of
$SO(3)_H$ as well \cite{bb,Babu:2014sga}.  In this case, although the soft masses of all members of the $SO(3)$ triplets would be degenerate
at the GUT scale, there is significant mixing between the third family and ceratin vector-like families of GUT scale mass that
generates a large top quark mass.  As a result, the effective low energy SUSY breaking Lagrangian would have a common mass for the
first two family sfermions that is different from that of the third family.  Thus, both $SU(2)$ and $SO(3)$ would lead to the parameter
set of Eq. (\ref{para}) for low energy phenomenology.

The spectrum of sMSSM can also follow from a non-Abelian discrete flavor symmetry such as
$S_3$ and $A_4$ \cite{Babu:2014sga}.  We envision such symmetries to have a fundamental gauge origin and note that in string theory constructions such
non-Abelian symmetries often emerge.  In this case there is no issue with the $D$-term, since discrete symmetries do not have
associated $D$-terms.  $S_3$, the permutation group of three letters, which is the simplest non-Abelian symmetry,
admits a {\bf 2+1} family assignment.  $A_4$, the symmetry group of a regular tetrahedron, which is the simplest group with a triplet
representation, admits a {\bf 3} assignment of families. Realistic fermion mass generation and symmetry breaking mechanism with
these symmetry groups have been analyzed in Ref. \cite{Babu:2014sga}, which all yield the spectrum of sMSSM.  The case of $S_3$ symmetry
is similar to the $SU(2)_H$ model, while the case of $A_4$ symmetry resembles the $SO(3)_H$ model.

sMSSM  is a systematized approach which addresses and solves the $D$-term problem \cite{murayama} that generally exists in gauged family symmetry models \cite{gauge} by auxiliary symmetries. Non-Abelian discrete family symmetries have  been used in the literature to address the SUSY flavor violation problem \cite{discrete1,discrete2}, but typically the low energy theory is not the MSSM.  In the sMSSM, on the other hand,
the low energy theory is the MSSM with the parameter set relevant for SUSY breaking given as in Eq. (\ref{para}).

We conclude this section with a brief description of one model based on $SU(2)_H$ flavor symmetry that yields sMSSM at low energies \cite{Babu:2014sga}.
The three families
are assigned under $SO(10) \times SU(2)_H$ as $({\bf 16},\,{\bf 2}) + ({\bf 16},\,{\bf 1})$, with the $({\bf 16},\,{\bf 1})$
identified as the third family.  We use the notation of $SO(10)$, but it is not required that the model be grand unified; the
only requirement is compatibility with a GUT symmetry such as $SO(10)$.  $SU(2)_H$ symmetry breaking is achieved
by introducing a pair of $({\bf 1},\,{\bf 2})$ Higgs fields, denoted as $\phi$ and $\overline{\phi}$, which acquire vacuum expectation
values (VEVs) of order the GUT scale, through a superpotential given by
\begin{equation}
W_{\rm sym} = \mu_\phi\, \phi \,\overline{\phi} + \kappa \,(\phi\, \overline{\phi})^2~.
\end{equation}
Here $\kappa$ is a parameter with inverse dimensions of mass, obtained by integrating a gauge singlet field, or arising from
quantum gravity effects. Including the $SU(2)_H$ $D$-terms, and soft SUSY breaking terms, one obtains from the minimization of the potential a condition
\begin{equation}
|u|^2 - |\overline{u}|^2 = \frac{2 (m^2_{\overline{\phi}} - m_\phi^2)}{g_H^2},
\label{split}
\end{equation}
where $\left \langle \phi \right\rangle = (0,\, u)^T$ and $\left \langle \overline{\phi} \right\rangle = (\overline{u},\,0)^T$,
and where $m^2_\phi$  and $m^2_{\overline{\phi}}$ are the soft squared masses of the $\phi$ and $\overline{\phi}$ fields respectively.
This condition yields a nonzero $D$-term, which would split the masses of the up and down--type members of all $SU(2)_H$ doublet sfermions,
and thus induce flavor violation (once CKM mixing is included) in meson--antimeson mixing, for example.
In Ref. \cite{Babu:2014sga} it was noted that this $D$-term problem can be avoided simply by imposing an interchange
symmetry $\phi \leftrightarrow \overline{\phi}$, which would set $m^2_\phi = m^2_{\overline{\phi}}$, and thus $|u|^2 = |\overline{u}|^2$.
Such an interchange symmetry is a subgroup of an anomaly free $SU(2)$ global symmetry which exists in the model with two doublets.

Realistic fermion masses are induced in the model through the Yukawa superpotential
\begin{equation}
W_{\rm Yuk} = {\bf 16}_3 {\bf 16}_3 {\bf 10}_H + {\bf 16}_i {\bf 16}_3 {\bf 10}_H \left(\frac{\phi_j +\overline{\phi}_j}{M_*}\right) \epsilon^{ij} +
{\bf 16}_i {\bf 16}_j \epsilon^{ij} {\bf 10}_H \left(\frac{{\bf 45}_H}{M_*}\right) + ...
\label{Yuk1}
\end{equation}
Here ellipsis  stands for higher order terms suppressed by more powers of $M_*$, which is presumable the Planck scale, much
larger than $|u|$ and $\left\langle {\bf 45}_H \right \rangle \sim M_{\rm GUT}$.   The coupling ${\bf 16}_i {\bf 16}_j \epsilon^{ij} {\bf 10}_H$
will not be allowed if the full $SO(10)$ symmetry is utilized, however the term shown with an additional ${\bf 45}_H$, used for GUT
symmetry breaking, would be allowed because of its antisymmetric property.  We see that only the third generation acquires a mass at the
renormalizable level, while the lighter family masses are suppressed by inverse powers such as $|u|/M_*$.  After some rotations, the
fermion mass matrices resulting from Eq. (\ref{Yuk1}) can be written in the form
\begin{eqnarray}
M_f = \left( \begin{matrix} 0 & c & 0 \\ -c & 0 & b \\ 0 & b' & a \end{matrix}
\label{Mf1}
\right)_f
\end{eqnarray}
for $f=u,\,d,\,\ell,\nu^D$, which fits the observed masses and mixings of quarks and leptons quite well \cite{discrete2}.  CP violation
can have a spontaneous origin in this context, which would make all SUSY breaking parameters real, and thus solve the SUSY CP problem
arising from limits on the electric dipole moments of the electron and the neutron.  The CKM phase can be still be of order one, if some
of the fields, such as the ${\bf 45}_H$ of Eq. (\ref{Yuk1}), acquire complex VEVs \cite{Babu:2014sga}.

Owing to the $SU(2)_H$ flavor symmetry, the soft masses of the scalars in the $({\bf 16}, \,{\bf 2})$ multiplet are all the same (denoted as
$m_{1,2}$), while members of the $({\bf 16},\,{\bf 1})$ would have a common mass that is different (denoted as $m_3$).  The gaugino masses
are unified because of the $SO(10)$ symmetry.  There is no reason for the soft masses of the MSSM Higgs doublets $H_u$ and $H_d$ to be
equal to $m_{1,2}$ or $m_3$, as these fields belong to different representations of $SO(10)$ such as {\bf 10} and {\bf 16}.  These two
Higgs soft masses have been traded in Eq. (\ref{para}) with $\mu$ and $m_A$.  Finally, in the sMSSM framework it is not required that the
trilinear $A$-terms be proportional to the respective Yukawa couplings.  Nevertheless, these $A$-terms would exhibit the same hierarchy
as the Yukawa couplings, and non-proportionality does not result in excessive SUSY induced flavor violation.  For low energy collider
phenomenology, only the third family $A$-terms are relevant, which we denote as $A_0$ at the GUT scale.  In a more general setting this
$A_0$ can break into $A_0^t$, $A_0^b$ and $A_0^\tau$, which need not be all the same.  Such a difference will be relevant only for the
case of large $\tan\beta$. In our analysis we define $A_0 = A_t^0=A_b^0=
A_\tau^0$, which is realized in at least some versions of sMSSM.

\section{\label{g-2}The Muon Anomalous Magnetic Moment in sMSSM}

The leading  contribution from low scale supersymmetry  to the muon anomalous magnetic moment, applicable to sMSSM,  is given by \cite{Moroi:1995yh, Martin:2001st}:

\bea
\label{eqq1}
\Delta a_\mu &=&
\frac{\alpha \, m^2_\mu \, \mu\,M_{2} \tan\beta}{4\pi \sin^2\theta_W \, m_{\tilde{\mu}_{L}}^2}
\left[ \frac{f_{\chi}(M_{2}^2/m_{\tilde{\mu}_{L}}^2)-f_{\chi}(\mu^2/m_{\tilde{\mu}_{L}}^2)}{M_2^2-\mu^2} \right]
\nonumber\\
&+&
\frac{\alpha \, m^2_\mu \, \mu\,M_{1} \tan\beta}{4\pi \cos^2\theta_W \, (m_{\tilde{\mu}_{R}}^2 - m_{\tilde{\mu}_{L}}^2)}
\left[\frac{f_{N}(M^2_1/m_{\tilde{\mu}_{R}}^2)}{m_{\tilde{\mu}_{R}}^2} - \frac{f_{N}(M^2_1/m_{\tilde{\mu}_{L}}^2)}{m_{\tilde{\mu}_{L}}^2}\right] \,,
\eea
where  $\alpha$ is the fine-structure constant, $m_\mu$ is the muon mass, $\mu$ denotes  the bilinear Higgs mixing term, and  $\tan\beta$ is the ratio of the vacuum expectation values (VEVs) of MSSM Higgs doublets. $M_1$ and $M_2$ denote the $U(1)_Y$ and $SU(2)$ gaugino masses respectively, $\theta_W$  is the weak mixing angle, and $m_{\tilde{\mu}_{L}}$,  $m_{\tilde{\mu}_{R}}$ are left and right handed smuon {masses}.
The loop functions are defined as follows:
\bea
f_{\chi}(x) &=& \frac{x^2 - 4x + 3 + 2\ln x}{(1-x)^3}~,\qquad ~f_{\chi}(1)=-2/3, \\
f_{N}(x) &=& \frac{ x^2 -1- 2x\ln x}{(1-x)^3}\,,\qquad\qquad f_{N}(1) = -1/3 \, .
\label{eqq2}
\eea
The first term in Eq. (\ref{eqq1}) stands for the dominant contribution arising  from the one loop diagram with Higgsino-wino exchange, while the
second term describes contributions  from the bino-smuon loop.
As the Higgsino mass parameter $\mu$ increases, the first term decreases in Eq. (\ref{eqq1}) and the   second term becomes dominant. {On the other  hand, the smuon  need to be light, $O$(100 GeV), in both cases} in order to make a  {sizeable} contribution to muon $g-2$.  Note that due to decoupling,  the formulae  will eventually fail to be accurate for large values of $\mu\tan\beta$.
The  Eq. (\ref{eqq1})
does not contain the trilinear SSB term $A_{\mu}$, since it is assumed that $A_{\mu}<\mu\tan\beta$.
From  Eq. (\ref{eqq1}),  the parameter set
\begin{equation}
\{M_1, \, M_2, \, \mu,\, \tan\beta,  m_{\tilde{\mu}_{L}}, \, m_{\tilde{\mu}_{R}}\},
\label{eqq3}
\end{equation}
is relevant at low energies for the  muon $g-2$  calculation. Since the gaugino masses are universal at the GUT scale in sMSSM, and
the sfermions of the first two families have a common mass,
we have $M_2\approx 2 M_1$ at low scale due to  renormalization group equation (RGE) running.  On the other  hand,  in order to have {sizeable} contribution to muon $g-2$ from supersymmetry, the  gauginos should   be sufficiently  light. Because of relatively small values of bino and wino masses  we can assume that
  $m_{\tilde{\mu}_{L}} \approx m_{\tilde{\mu}_{R}}$. With these constraints the number of independent {parameters} for the $g-2$ calculation  can be reduced to four:
\begin{equation}
\{ M_1,  \, \mu,\, \tan\beta,   \, m_{\tilde{\mu}_{R}}\}.
\label{eqq4}
\end{equation}
We pay special attention to these parameters, which are functions of the seven fundamental parameters shown in Eq. (\ref{para}) in sMSSM.


\section{ Scanning Procedure, Parameter Space and Experimental Constraints\label{constraintsSection}}

We employ the ISAJET~{7.84} package~\cite{ISAJET}  to perform random
scans over the fundamental parameter space of sMSSM as shown in Eq. (\ref{para}).
In this package, the weak scale values of gauge and third generation Yukawa
couplings are evolved to $M_{\rm GUT}$ via the MSSM RGEs in the $\overline{DR}$ regularization scheme.
We do not strictly enforce the unification condition $g_3=g_1=g_2$ at $M_{\rm
GUT}$, since a few percent deviation from unification can be
assigned to unknown GUT-scale threshold
corrections~\cite{Hisano:1992jj}.
The deviation between $g_1=g_2$ and $g_3$ at $M_{\rm GUT}$ is no
worse than $3-4\%$.
For simplicity,  we do not include the Dirac neutrino Yukawa coupling
in the RGEs, whose contribution is expected to be small.

The various boundary conditions are imposed at
$M_{\rm GUT}$ and all the SSB
parameters, along with the gauge and Yukawa couplings, are evolved
back to the weak scale $M_{\rm Z}$.
In the evolution of Yukawa couplings the SUSY threshold
corrections~\cite{Pierce:1996zz} are taken into account at the
common scale $M_{\rm SUSY}= \sqrt{m_{{\tilde t}_L}m_{{\tilde t}_R}}$,
where $m_{{\tilde t}_L}$ and $m_{{\tilde t}_R}$
denote the masses of the third generation left and right-handed stop quarks.
The entire parameter set is iteratively run between $M_{\rm Z}$ and $M_{\rm
GUT}$ using the full 2-loop RGEs until a stable solution is
obtained. To better account for leading-log corrections, one-loop
step-beta functions are adopted for the gauge and Yukawa couplings, and
the SSB parameters $m_i$ are extracted from RGEs at multiple scales
$m_i=m_i(m_i)$. The RGE-improved 1-loop effective potential is
minimized at $M_{\rm SUSY}$, which effectively
accounts for the leading 2-loop corrections. Full 1-loop radiative
corrections are incorporated for all sparticle masses.

An approximate error of around 2 GeV \cite{Degrassi:2002fi} in the estimate of the Higgs boson mass
largely arises from theoretical uncertainties in the  {calculation} of the minimum of the
scalar potential, and to a lesser extent from experimental uncertainties in the values
for $m_t$ and $\alpha_s$.

An important constraint on the parameter space arises from limits on the cosmological abundance of stable charged
particles  \cite{Nakamura:2010zzi}. This excludes regions in the parameter space
where a charged SUSY particle  becomes
the lightest supersymmetric particle (LSP). We accept only those
solutions for which one of the neutralinos is the LSP and saturates
the WMAP  bound on relic dark matter abundance.

We have performed random scans for the following parameter range:
\begin{align}
0 \leq  m_{1,2}  \leq 3\, \rm{TeV} \nonumber  \\
0 \leq  m_{3}  \leq 3\, \rm{TeV} \nonumber  \\
0 \leq  M_{1/2}  \leq 3\, \rm{TeV} \nonumber  \\
-5\, \rm{TeV} \leq A_{0}  \leq 5\,  \rm{TeV} \nonumber  \\
-3\leq A_{0}/m_{3} \leq 3 \nonumber \\
2 \leq  \rm {tan}\beta  \leq 60 \nonumber \\
0 \leq  \mu  \leq 3\, \rm{TeV} \nonumber \\
0 \leq  m_{A}  \leq 3\, \rm{TeV} \nonumber \\
\mu > 0.
\label{parameterRange}
\end{align}
{Here $m_{1,2}$ is the SSB scalar mass parameters for the first two generations, while $m_{3}$ is for the third generation. $M_{1/2}$ is the SSB gaugino mass, and $A_{0}$ is the SSB trilinear scalar interaction coupling. The parameters $\mu$ and $m_{A}$ are  bilinear Higgs mixing term and mass of the CP-odd Higgs boson respectively. In contrast to the other parameters, the values for $\mu$ and $m_{A}$  are set at low scale. We} make  $m_t = 173.3\, {\rm GeV}$  \cite{:1900yx}, and we show that our results are not
too sensitive to one or two sigma variations from this central value   \cite{Gogoladze:2011db}.
Note that $m_b(m_Z)=2.83$ GeV, which is hard-coded into ISAJET. The choice of the sgn($\mu$) to be positive is dictated by the desire to
explain the muon $g-2$ anomaly.  All SUSY breaking parameters are restricted to lie below 3 TeV (except for $A_0$ which is allowed to
be somewhat larger), which would make the fine tuning in the
Higgs mass relatively mild.  Since most of the SUSY particles have masse below about 4 TeV, essentially all particles are within reach
of the LHC.

\begin{figure}[]
\label{ffigure1}
\subfigure{\includegraphics[scale=1.0]{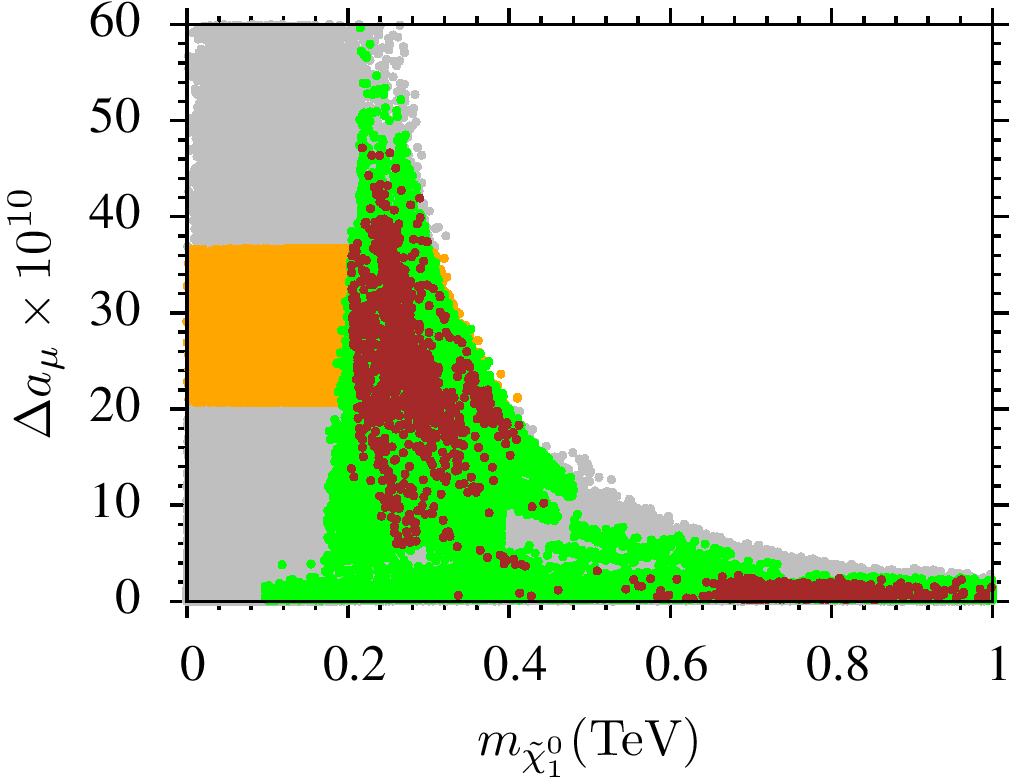}}
\subfigure{\includegraphics[scale=1.0]{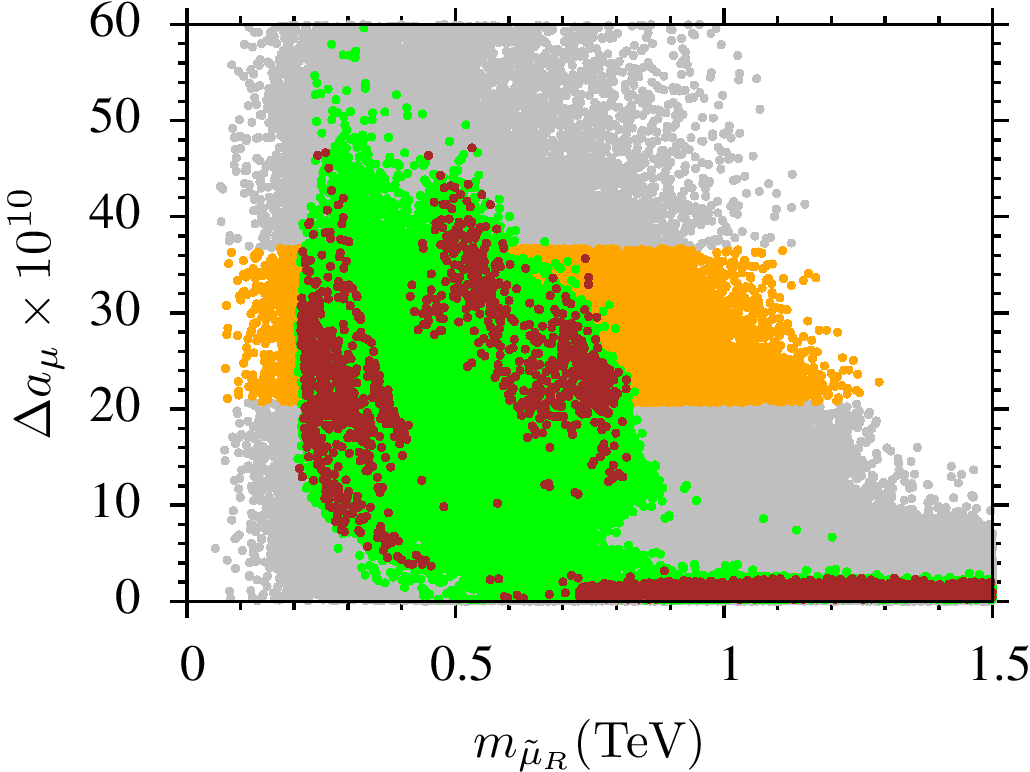}} \\
\subfigure{\includegraphics[scale=1.0]{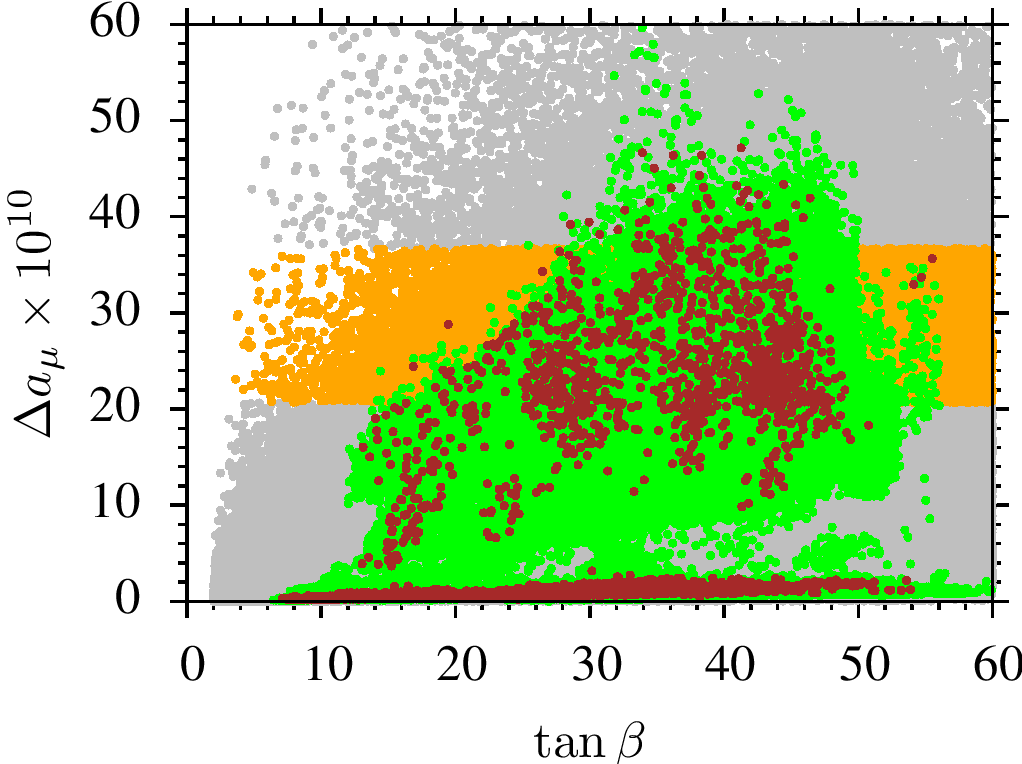}}
\subfigure{\includegraphics[scale=1.0]{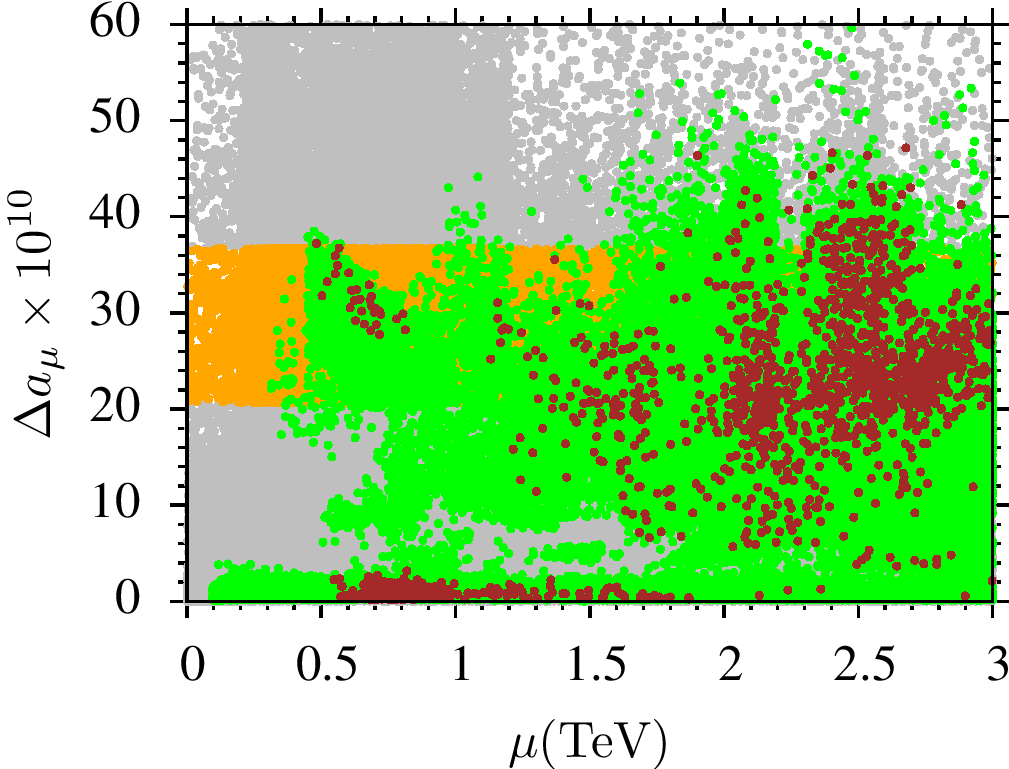}}
\caption{Plots in the $\Delta a_{\mu}-m_{\tilde{\chi}^{0}_{1}} $, $\Delta a_{\mu}-m_{\tilde{\mu}_{R}} $, $\Delta a_{\mu}-\tan\beta $, $\Delta a_{\mu}-\mu $ planes.
Gray points are consistent with REWSB  and neutralino LSP. Yellow points represent values of $\Delta a_{\mu}$ that would bring
theory and experiment to within $1\sigma$. Green points form a subset of gray points and satisfy  sparticle mass and $B$-physics constraints described on Table \ref{table1}. In addition these
points satisfy the lightest CP-even Higgs mass range  $123\, {\rm GeV} \leq m_h \leq127$ \,{\rm GeV}. Brown points belong to a subset of green points and satisfy the WMAP bound ($5\sigma$) on neutralino dark matter abundance.
}
\end{figure}

In scanning the parameter space, we employ the Metropolis-Hastings
algorithm as described in \cite{Belanger:2009ti}. The data points
collected all satisfy
the requirement of radiative electroweak symmetry breaking (REWSB),
with the neutralino {being the LSP in each case}. After collecting the data, we impose
the mass bounds on  the particles \cite{Nakamura:2010zzi} and use the
IsaTools package~\cite{Baer:2002fv}
to implement the various phenomenological constraints. We successively apply the  experimental constraints presented in Table 1 on the data that
we acquire from ISAJET:

\begin{table}[h]\centering
\begin{tabular}{rlc}
$   123\, {\rm GeV} \leq m_h \leq127$ \,{\rm GeV}~~~&\cite{:2012gk,:2012gu}&
\\
$0.8\times 10^{-9} \leq{\rm BR}(B_s \rightarrow \mu^+ \mu^-)
  \leq 6.2 \times10^{-9} \;(2\sigma)$~~~&\cite{BsMuMu}&
\\
$2.99 \times 10^{-4} \leq
  {\rm BR}(b \rightarrow s \gamma)
  \leq 3.87 \times 10^{-4} \; (2\sigma)$~~~&\cite{Amhis:2012bh}&
\\
$0.15 \leq \frac{
 {\rm BR}(B_u\rightarrow\tau \nu_{\tau})_{\rm MSSM}}
 {{\rm BR}(B_u\rightarrow \tau \nu_{\tau})_{\rm SM}}
        \leq 2.41 \; (3\sigma)$~~~&\cite{Asner:2010qj}&
\\
$ 0.0913 \leq \Omega_{\rm CDM}h^2  \leq 0.1363    \; (5\sigma)$~~~&\cite{WMAP9}&
\end{tabular}
\caption{Phenomenological constraints implemented in our study.}
\label{table1}
\end{table}

\section{Results \label{spectroscopySection}}

\begin{figure}[]
\label{ff2}
\centering
\includegraphics[scale=1.05]{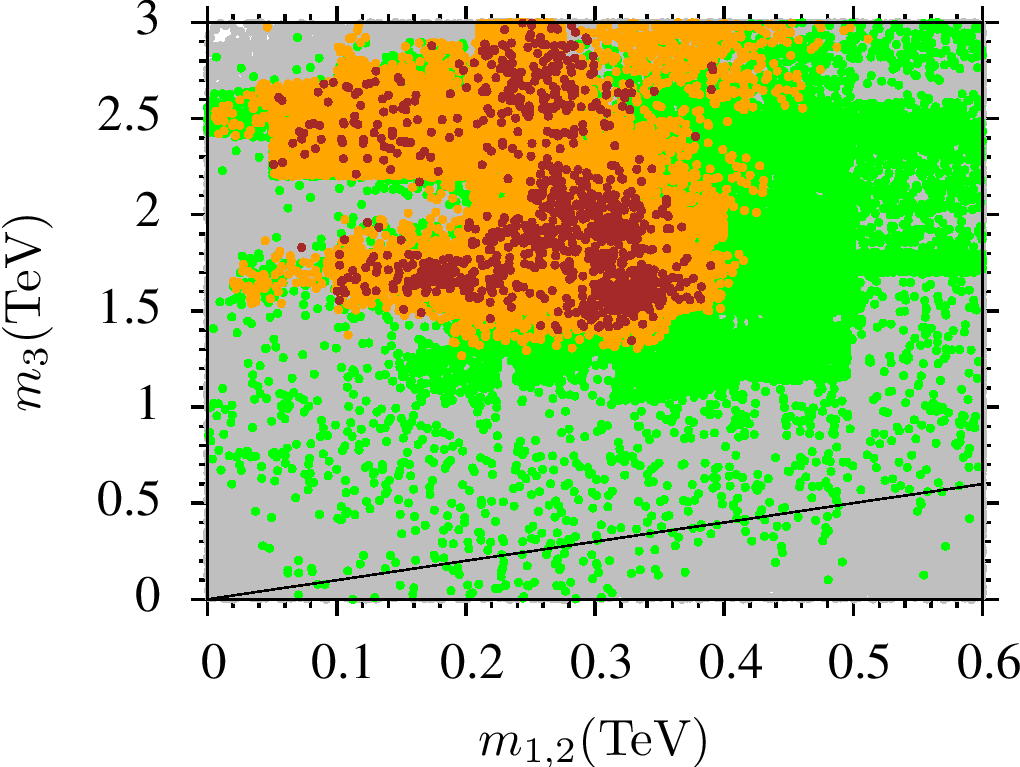}
\caption{Plots in the $m_{3}-m_{1,2} $ plane. The color coding is the same as Figure 1,
but in this case yellow points are a subset of green points and brown points belong to a subset of yellow.
The unit slope line is to guide the eye. }
\end{figure}

\begin{figure}[]
\label{ff3}
\subfigure{\includegraphics[scale=1.05]{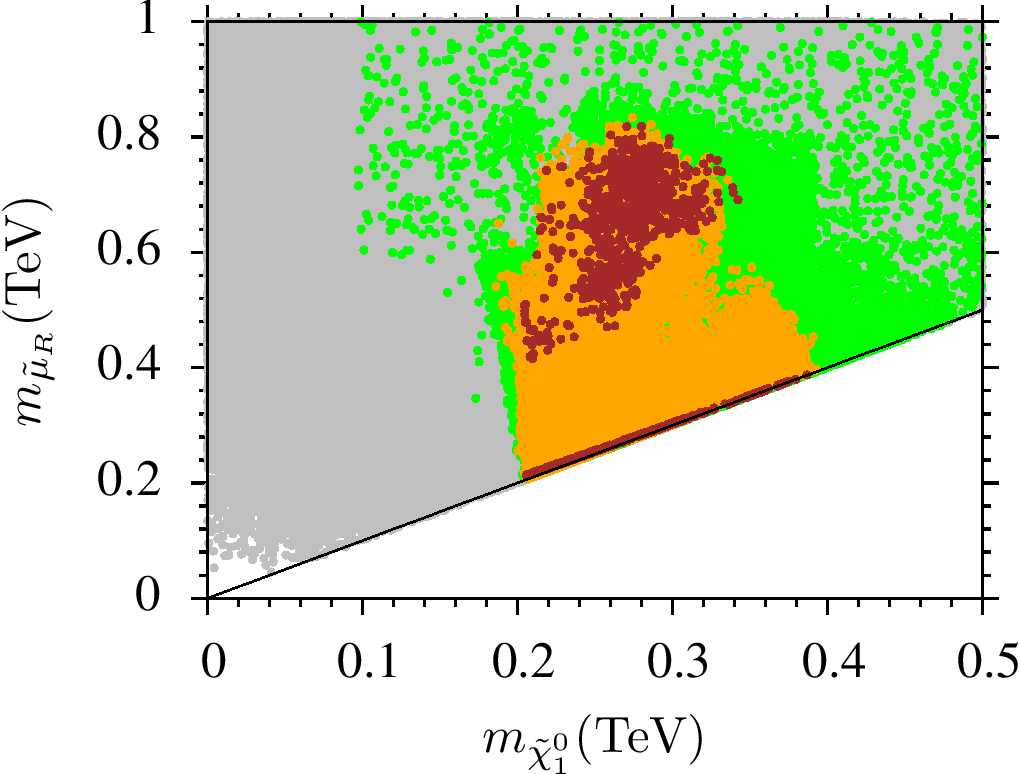}}
\subfigure{\includegraphics[scale=1.05]{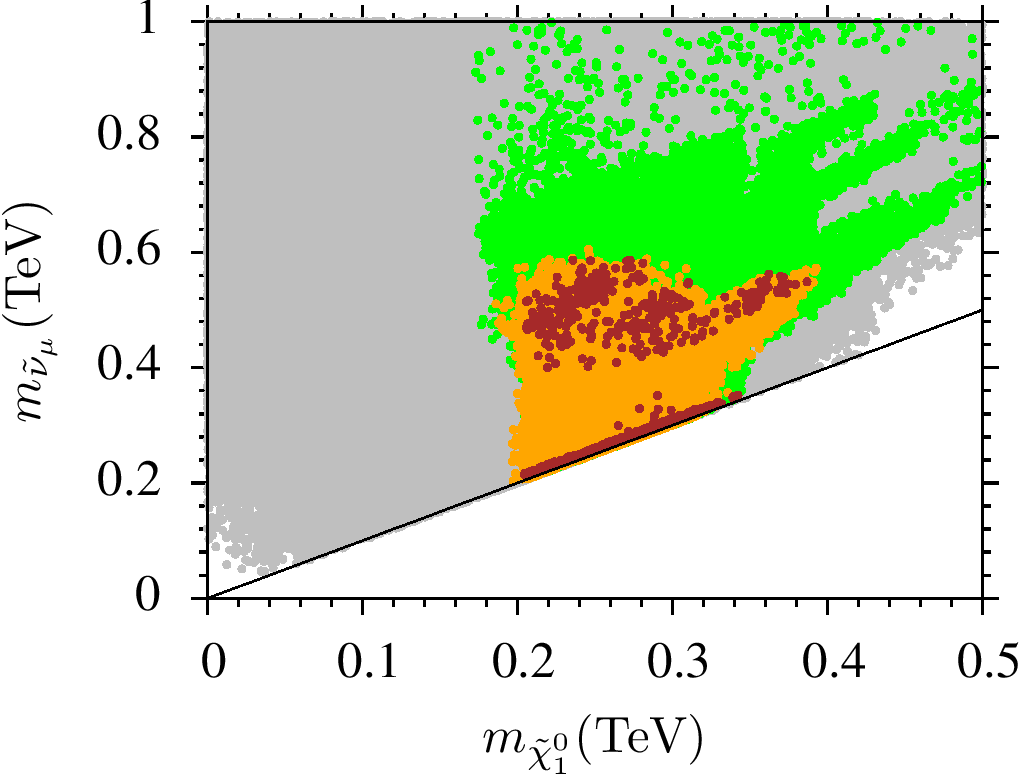}} \\
\subfigure{\includegraphics[scale=1.05]{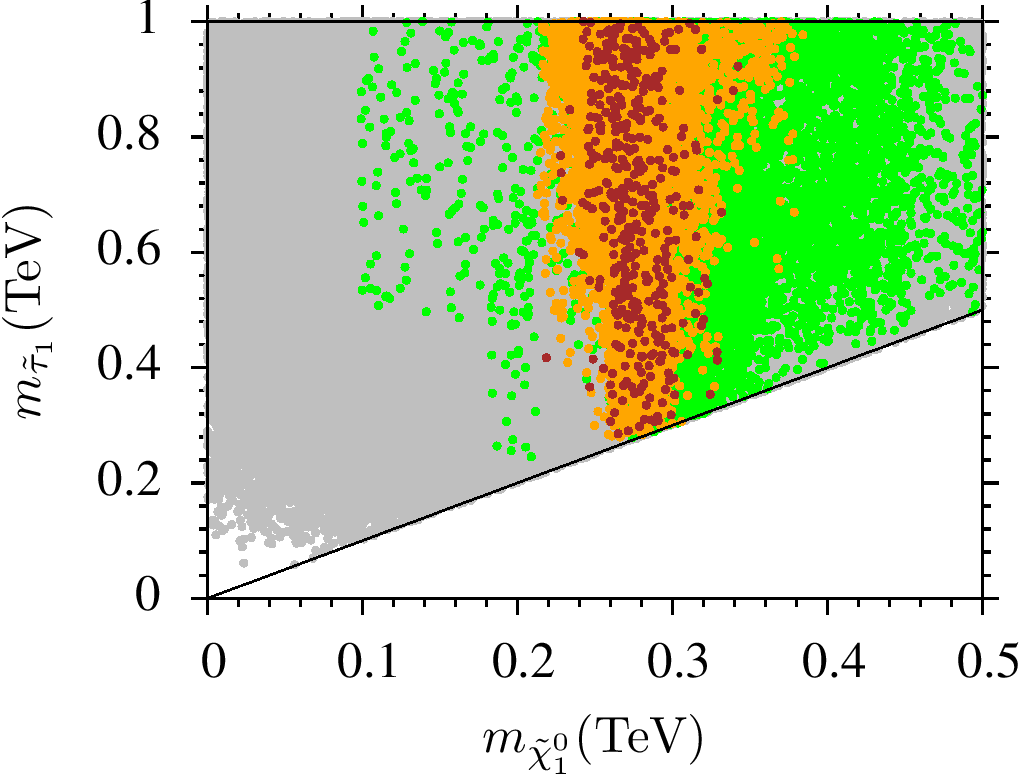}}
\subfigure{\includegraphics[scale=1.05]{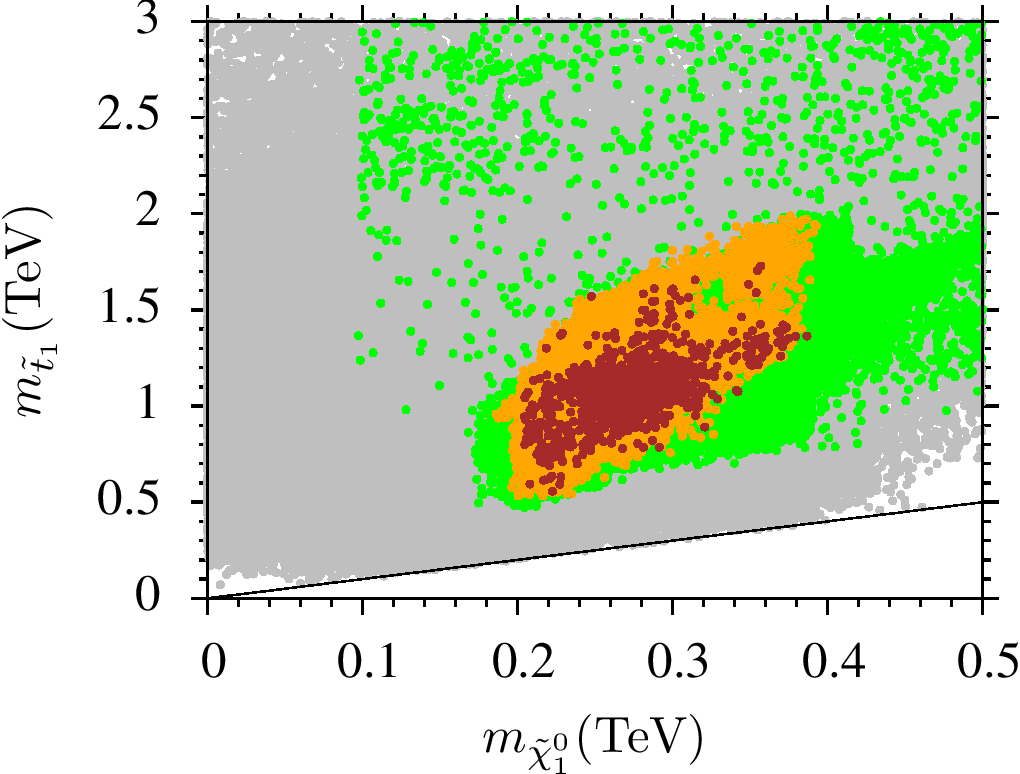}}
\caption{Plots in the  $m_{\tilde{\mu}_{R}}-m_{\tilde{\chi}^{0}_{1}} $, $m_{\tilde{\nu}_{1,2}}-m_{\tilde{\chi}^{0}_{1}} $, $m_{\tilde{\tau}_{1}}-m_{\tilde{\chi}^{0}_{1}} $, $m_{\tilde{t}}-m_{\tilde{\chi}^{0}_{1}} $ planes. The color coding is the same as Figure 2 except that the mass bound on stop is not applied in $m_{\tilde{t}}-m_{\tilde{\chi}^{0}_{1}} $.  }
\end{figure}


\begin{figure}[]
\label{ff4}
\subfigure{\includegraphics[scale=1.05]{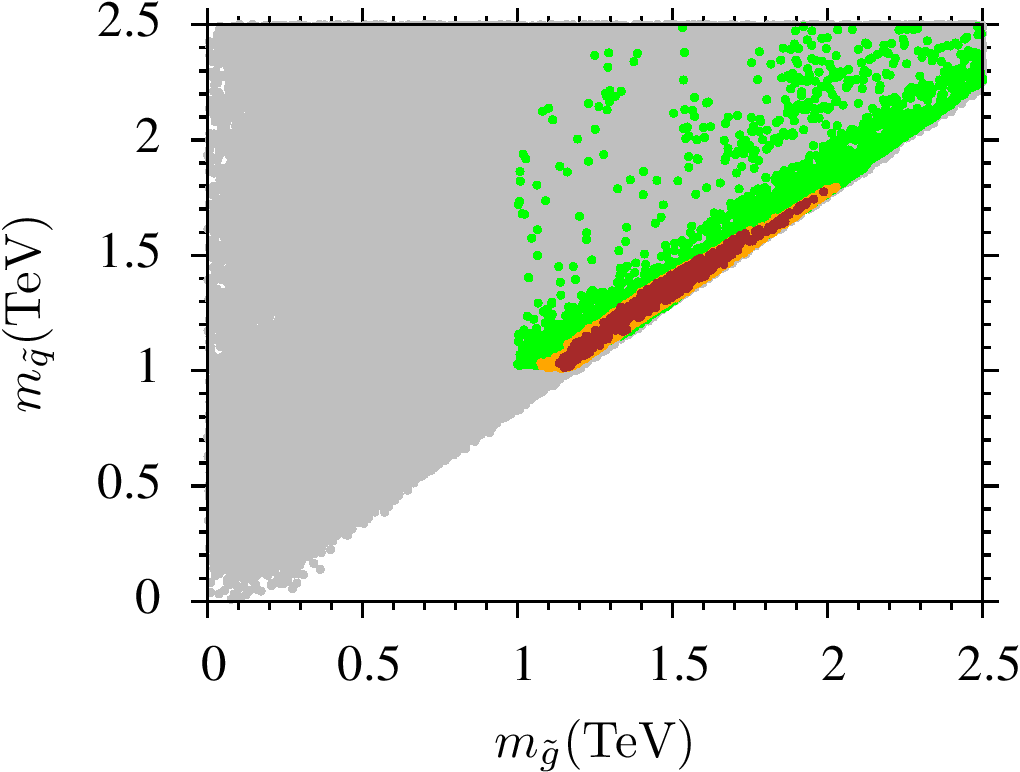}}
\subfigure{\includegraphics[scale=1.05]{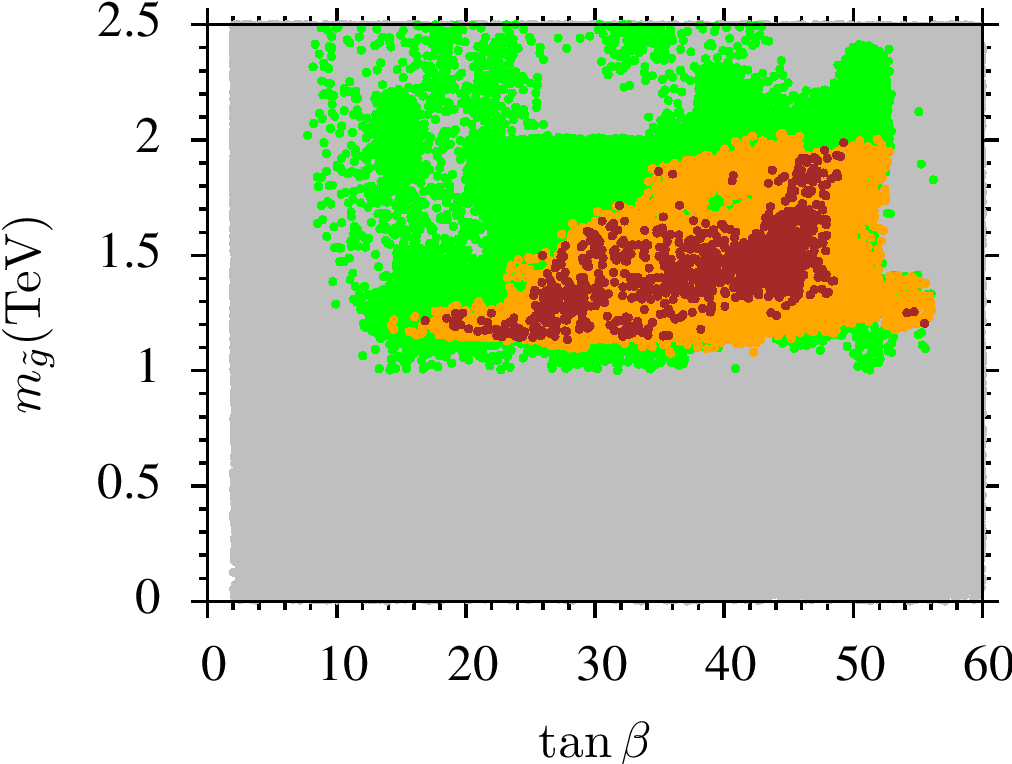}}
\caption{Plots in the $m_{\tilde{q}}-m_{\tilde{g}} $ and  $ m_{\tilde{g}}-\tan\beta $  planes. Color coding is the same as Figure 2.}
\end{figure}

\begin{table}[h!]\hspace{-1.0cm}
\centering
\begin{tabular}{|c|cccc|}
\hline
\hline
                 & Point 1 & Point 2 & Point 3 & Point 4 \\

\hline
$m_{1,2}$        & 222 & 302 & 282 & 244 \\
$m_{3}$          & 2862 & 1760 & 1678  & 2671\\
$M_{1/2} $       & 545.6 & 494 & 692 & 754 \\
$\tan\beta$      & 35.4 & 20.9 & 44.4 & 46.1 \\
$A_0/m_{3}$      & -1.54 & -2.24 & -2.65 & -2.18\\
$\mu$            & 503.1 & 2179 & 2676 & 2895 \\
$m_{A}$          & 2891 & 1648 & 2749 & 2972 \\
$m_t$            & 173.3  & 173.3 & 173.3 & 173.3 \\
\hline
& & & & \\
$\Delta a_{\mu}$  & $ \mathbf{ 31.8\times 10^{-10} } $ & $ \mathbf{24.3\times 10^{-10} } $ & $ \mathbf{ 22.5\times 10^{-10} } $ & $ \mathbf{23.1 \times 10^{-10} } $ \\ & & & & \\

\hline
$m_h$            & \textbf{123.2} & \textbf{124.1} & \textbf{124.6} & \textbf{125.2} \\
$m_H$            & 2910 & 1658 & 2767  & 2991 \\
$m_A$            & 2891 & 1648 & 2749 & 2972 \\
$m_{H^{\pm}}$    & 2911 & 1661 & 2768 & 2993 \\

\hline
$m_{\tilde{\chi}^0_{1,2}}$
                 & {\color{red} 232},420.7 & {\color{red} 211}, 410 & {\color{red} 299}, 573 & {\color{red} 330}, 631 \\

$m_{\tilde{\chi}^0_{3,4}}$
                 & 514.2, 548 & 2164, 2164 & 2658, 2658 & 2874, 2875 \\

$m_{\tilde{\chi}^{\pm}_{1,2}}$
                 & 423.5, 546.5 & 411, 2169 & 574, 2659 & 633, 2877 \\

$m_{\tilde{g}}$  & 1290 & 1171 & 1579 & 1724 \\
\hline $m_{ \tilde{u}_{L,R}}$
                 & 1137, 1041 & 1465, 1298  & 1399, 1218 & 1561, 1401 \\
$m_{\tilde{t}_{1,2}}$
                 & 1066, 1960 & 896, 1553 & 1019, 1597 & 1267, 2030 \\
\hline $m_{ \tilde{d}_{L,R}}$
                 & 1140, 1117.5 & 1069, 1022 & 1468, 1431 & 1563, 1521 \\
$m_{\tilde{b}_{1,2}}$
                 & 1976, 2466 & 1532, 1892 & 1545, 1755 & 2014, 2354 \\
\hline
$m_{\tilde{\nu}_{1,2}}$
                 & {\color{red} 244} & 473 & 326 & {\color{red}340}\\
$m_{\tilde{\nu}_{3}}$
                 &  2541 & 1724 & 1146 & 2021\\
\hline
$m_{ \tilde{e}_{L,R}}$
                & 319, 474 & 491, {\color{red} 218} & 355, 706 & 387, 687  \\
$m_{\tilde{\tau}_{1,2}}$
                & 2195, 2546  & 1581, 1731 & {\color{red} 318}, 1159  & 1109, 2025\\
\hline

$\sigma_{SI}({\rm pb})$
                & $0.35\times 10^{-9}$ & $ 0.53\times 10^{-11} $ & $ 0.36\times 10^{-11} $ & $0.13\times 10^{-11} $ \\

$\sigma_{SD}({\rm pb})$
                & $0.19\times 10^{-5}$ &$ 0.44\times 10^{-7} $ & $ 0.43\times 10^{-8} $ & $0.32\times 10^{-8} $ \\

$\Omega_{CDM}h^{2}$
                &  0.11 & 0.11 & 0.12 & 0.11\\

\hline
\hline
\end{tabular}
\caption{ Four benchmark points satisfying all phenomenological constraints including muon $g-2$ in sMSSM.
All the masses are in units of GeV.
All this points  are chosen to  satisfy
 the  constraints described in Section 3.
 The points 1-4 respectively correspond to  muon  sneutrino, smuon, stau and  muon  sneutrino coannihilation  channels. }
\end{table}

We next {present the  results} of the scan over the parameter space listed in Eq. (\ref{parameterRange}).
In Figure 1  we show the results in the $\Delta a_{\mu}-m_{\tilde{\chi}^{0}_{1}} $, $\Delta a_{\mu}-m_{\tilde{\mu}_{R}} $, $\Delta a_{\mu}-\tan\beta $, $\Delta a_{\mu}-\mu $ planes. Gray points are consistent with REWSB  and neutralino LSP. Yellow points represent $\Delta a_{\mu}$ values
which would bring theory and experiment within  $1\sigma$. Green points form  a subset of gray points and satisfy the sparticle mass bounds  and $B$-physics constraints described in Table \ref{table1}. In addition,  the lightest CP-even Higgs mass range  $123\, {\rm GeV} \leq m_h \leq 127$ \,{\rm GeV} {is applied}.   Brown points belong to a subset of green points and satisfy the WMAP bound ($5\sigma$) on the neutralino dark matter abundance.

 In the  $\Delta a_{\mu}-m_{\tilde{\chi}^{0}_{1}} $  plane of Figure 1,  we show  that muon $g-2$
prefers  relatively light gauginos in the SUSY spectrum for  $\Delta a_{\mu}$ to be {large  enough to} explain the discrepancy between theory and experiment. The brown points  belong to a subset of green points and satisfy the WMAP bound ($5\sigma$) on  neutralino dark matter  abundance. {We will consider later on how to obtain the correct relic abundance of neutralino dark matter in this model. The lower  bound on the  neutralino mass arises  mostly from the  current gluino mass bound,  and there is a   sharp upper bound on the former, of about 2 TeV, if we want to have $\Delta a_{\mu}$ within a $1\sigma$ deviation.

From the $\Delta a_{\mu}-m_{\tilde{\mu}_{R}} $ {panel,} we see that in order to stay within a  $1\sigma$ range   of muon $g-2$ and comply with all the {constraints} listed in Section \ref{constraintsSection}, the  smuon mass should lie in the range $200\, {\rm GeV} \lesssim m_{\tilde{\mu}_{R}} \lesssim 800\, {\rm GeV}$.

The results in the $\Delta a_{\mu}-\tan\beta $ plane show  that    it is hard to have substantial contribution to muon $g-2$
 if $\tan\beta \lesssim 14$. The interval $30 \lesssim \tan\beta \lesssim 50$ is preferred from the   muon $g-2$ point of view, which is also a desirable range for $t-b-\tau$ Yukawa coupling unification as well \cite{yukawaUn, Ajaib:2013zha}.

It is interesting to see {from the } $\Delta a_{\mu}-\mu $ plane that there exist large $\mu$ solutions,  which means that in this case we have significant contribution from the bino-smuon loop. It has bees  shown \cite{Endo:2013lva} that if bino and smuons are dominant {contributors} to the muon $g-2$, {the} corresponding parameter space  for sleptons can be tested at the LHC and ILC. The $\Delta a_{\mu}-\mu $ plane also {shows} the  possibility of smaller $\mu$ values   consistent with desirable values for muon $g-2$. Small values of $\mu$-term may make ``the little hierarchy" problem less  severe.

It is interesting to show the amount of  mass splitting  necessary between the third and first {two-family} sfermion SSB masses in order to satisfy all current phenomenological {constraints} including muon $g-2$.  We present our results in  the $m_{3}-m_{1,2} $ plane in Figure \ref{ff2}. The color coding is the same as Figure 1 but in this case the yellow points are a subset of the  green points, and the brown points belong to a subset of yellow. The unit slope line is to guide the eye. As we see, the yellow points are sufficiently above the unit line, and  we need to have $m_3/m_{1,2}>4$. The splitting {becomes larger } ($m_3/m_{1,2}>10$) as $\tan\beta$  decreases.

In Figure 3 we show the relic density channels consistent with muon ($g-2$) in the  $m_{\tilde{\mu}_{R}}-m_{\tilde{\chi}^{0}_{1}} $, $m_{\tilde{\nu}_{\mu}}-m_{\tilde{\chi}^{0}_{1}} $, $m_{\tilde{\tau}_{1}}-m_{\tilde{\chi}^{0}_{1}} $, $m_{\tilde{t}}-m_{\tilde{\chi}^{0}_{1}} $ planes.
We  see that a variety of coannihilation  scenarios are
compatible with muon $g-2$ and neutralino  dark matter.
In the  $m_{\tilde{\mu}_{R}}-m_{\tilde{\chi}^{0}_{1}} $
plane in Figure 3, we draw the unit slope
line which indicates the presence of smuon-neutralino coannihilation scenario.
From the $m_{\tilde{\nu}_{\mu}}-m_{\tilde{\chi}^{0}_{1}} $ and  $m_{\tilde{\tau}_{1}}-m_{\tilde{\chi}^{0}_{1}} $ planes we see that is is also possible to realize stau and muon sneutrino coannihilation scenarios.

The results in the $m_{\tilde{\tau}_{1}}-m_{\tilde{\chi}^{0}_{1}}$ plane show  that it is hard to realize  stop coannihilation scenario in this framework. The    stop in this scenario can be as light as 500 GeV and {cannot} be heavier than 2 TeV.
We expect that the $A$-funnel scenario is
also consistent with muon $g-2$, although we have not found it, perhaps due to lack of statistics.

Figure 4 shows plots in the $m_{\tilde{q}}-m_{\tilde{g}} $, $ m_{\tilde{q}}-\tan\beta $, $ m_{\tilde{g}}-\tan\beta $ and $ m_{\tilde{\mu}_{R}}-\tan\beta $ planes, with color coding  the same as in  Figure 2.
The $m_{\tilde{q}}-m_{\tilde{g}} $  plane shows that imposing $1\sigma$ deviation from the measured muon $g-2$ requires the first and second generation squark masses to be less {than} 2 TeV, which can be  tested in the  on upcoming LHC second run.
If the  bound $m_{\tilde{g}} \gtrsim  1.4~{\rm TeV}~ ({\rm for}~ m_{\tilde{g}}\sim m_{\tilde{q}})$,  observed from an {analysis} based on the  cMSSM parameter space, is confirmed for the case of general low scale SUSY, then $\tan\beta \lesssim 30$ will be excluded in this scenario.

Finally, in Table 2 we present four  characteristic benchmark points which summarize the
salient features of this model. For these  points  the $g-2$ constraints as well as sparticle mass and $B$-physics constraints described in Section \ref{constraintsSection} are satisfied.
 The points 1-4 respectively correspond to  muon  sneutrino, smuon, stau and  muon  sneutrino coannihilation  channels.  Point 1  depicts a solution with a relatively low value of $\mu$ and accordingly it has
  relatively large neutralino-nucleon spin-independent and
spin dependent  cross section, which can be tested at the upcoming
SuperCDMS, XENON 1T and IceCube DeepCore  experiments.
 Point 4 displays a solution with heavy gluino and squarks of the first two families.


\section{ Conclusion \label{conclusions}}

We have explored the sparticle and Higgs phenomenology of the flavor symmetry--based MSSM framework, referred to here as sMSSM.
Such models are motivated by a grand unified symmetry such as $SO(10)$ along with a non-Abelian flavor symmetry that suppresses SUSY flavor
violation.  The SUSY breaking Lagrangian in sMSSM is the most general consistent with these two symmetries.
Explicit ultra-violet complete models that generate sMSSM spectrum at low energies have been presented.  These include models
based on $SU(2)$ and $SO(3)$ gauged flavor symmetries, as well as those based on non-Abelian discrete symmetries such as $S_3$ and $A_4$.
The SUSY phenomenology of these models is described by seven parameters listed in Eq. (\ref{para}).
sMSSM contains three additional parameters compared to cMSSM. Specifically, the (common) soft mass
of the first two family sfermions is different from that of the third family.  This freedom helps us explain the muon $g-2$ anomaly, along
with the Higgs boson mass and the correct relic abundance of neutralino dark matter.  The parameter space is still rather restrictive, and
we have shown that the simultaneous explanation of these observables requires the mass of the gluino to be less than about 2 TeV, and the
mass of the first two family sleptons to be less than about 800 GeV.  The parameter $\tan\beta$ is preferred to be relatively large,
$\tan\beta > 15$.


\section*{Acknowledgments}
We would like to thank  Shabbar Raza    for useful discussions.
This work is supported in part by the DOE Grant Nos. de-sc0010108  (KSB) and DE-FG02-91ER40626 (IG and QS).
 This work used the Extreme Science and Engineering Discovery Environment (XSEDE), which is supported by the National Science
Foundation Grant Number OCI-1053575.
IG acknowledges support from the  Rustaveli
National Science Foundation  No. 31/98.
KSB and IG would like to thank CETUP* (Center for Theoretical Underground Physics and Related
Areas), supported by the US Department of Energy under Grant No. de-sc0010137 and by the
US National Science Foundation under Grant No. PHY-1342611, for its hospitality and partial
support during the 2013 Summer Program.  They also wish to thank Barbara Szczerbinska for providing
a stimulating atmosphere in Deadwood during the CETUP* 2013 program.


\end{document}